\documentclass{aa} 

\newcommand{\source}{GS~1826--24}
\newcommand{\inte}{\textsl{INTEGRAL}}
\newcommand{\xte}{\textsl{RXTE}}
\newcommand{\xmm}{\textsl{XMM-Newton}}

\newcommand{\swift}{\textsl{Swift}}

\def\ergcms{{\rm erg\,cm^{-2}\,s^{-1}}}
\def\cm2{{\rm cm^{-2}}}
\def\Msun{{\rm M_{\odot}}}

\usepackage{graphicx}
\usepackage[normalem]{ulem}
\usepackage{natbib}
\usepackage{textcomp} 

\usepackage{txfonts}
\begin{document}

   \title{Burst-induced coronal cooling in GS~1826--24}

 \subtitle{The clock wagging its tail}

   \author{C. S\'anchez-Fern\'andez\inst{1}
          \and
          J.~J.~E.~Kajava\inst{1,2} \and
         J.~Poutanen\inst{3,4,5} \and
          E.~Kuulkers\inst{6} \and
          V.~F. Suleimanov\inst{7,8,4}}

   \institute{European Space Astronomy Centre (ESA/ESAC), Science Operations Department, 28691 Villanueva de la Ca\~nada, Madrid, Spain\\
              \email{Celia.Sanchez@sciops.esa.int} 
              \and Finnish Centre for Astronomy with ESO (FINCA),   FI-20014 University of Turku, Finland
     \and Department of Physics and Astronomy,  FI-20014 University of Turku, Finland 
\and Space Research Institute of the Russian Academy of Sciences, Profsoyuznaya str. 84/32, 117997 Moscow, Russia 
\and Nordita, KTH Royal Institute of Technology and Stockholm University, Roslagstullsbacken 23, SE-10691 Stockholm, Sweden
     \and ESA/ESTEC, Keplerlaan 1, 2201 AZ Noordwijk, The Netherlands
     \and Institut f\"ur Astronomie und Astrophysik, Kepler Center for Astro and Particle Physics, Universit\"at T\"ubingen, Sand 1, 72076 T\"ubingen, Germany
    \and Astronomy Department, Kazan (Volga region) Federal University, Kremlyovskaya str. 18, 420008 Kazan, Russia}
   \date{Received xxx; accepted xxx}

 
\abstract
{Type I X-ray bursts in \source, and in several other systems, may induce cooling of the hot  inner accretion flow that surrounds the bursting neutron star. Given that \source\ remained persistently in the hard state over the period 2003--2008 and presented regular bursting properties, we stacked the spectra of the X-ray bursts detected by \inte\ (JEM-X and ISGRI) and \xmm\ (RGS)  during that period to study the effect of the burst photons on the properties of the Comptonizing medium.
The extended energy range provided by these instruments allows the simultaneous observation of the burst and persistent emission spectra. We detect an overall change in the shape of the persistent emission spectrum in response to the burst  photon shower. 
For the first time, we observe simultaneously a drop in the hard X-ray emission, together with a soft X-ray excess with respect to the burst blackbody emission.
The hard X-ray drop can be explained by burst-induced coronal cooling, while the bulk of the soft X-ray excess can be described by fitting the burst emission with an atmosphere model, instead of a simple blackbody model.
Traditionally, the persistent emission was assumed to be invariant during X-ray bursts, and more recently to change only in normalization but not in spectral shape; the observed change in the persistent emission level during X-ray bursts may thus trigger the revision of existing neutron star mass-radius constraints, as the derived values rely on the assumption that the persistent emission does not change during X-ray bursts. 
The traditional burst fitting technique leads to up to a 10\%\ overestimation of the bolometric burst flux in \source, which significantly hampers the comparisons of the KEPLER and MESA model  against this `textbook burster'.
}
   \keywords{stars: neutron -- X-rays: binaries -- X-rays: bursts -- accretion, accretion discs}

   \maketitle
%

\section{Introduction}

Type I X-ray bursts are thermonuclear explosions on the surface layers of weakly magnetized neutron stars (NSs) accreting mass from a low-mass companion (see reviews in \citealt{LvPT93,SB06}). 
The accreted hydrogen and/or helium accumulate steadily on top of the NS surface until they reach ignition temperatures and densities, and a thermonuclear runaway is triggered (e.g. \citealt{WT76,J78,WW81}). 
The sudden release of nuclear binding energy rapidly heats  the NS ocean and atmosphere, and within a few seconds increases the luminosity  to its peak value. 
The burst X-ray spectra are usually fit with a blackbody  ($kT$ peaking at 2--3\,keV; \citealt{SBB77}) that cools down during the burst decay.

Accreting NSs  are embedded in the accretion disc, and X-ray bursts are known to influence their surroundings in several ways.
In the hot flow paradigm  (e.g. \citealt{DGK07}) the inner accretion disc  is believed to puff up in the hard state to a geometrically thick and optically thin hot flow. 
The hot inner flow Compton up-scatters low-energy photons emitted by the truncated thin disc, the heated NS surface, and possibly  the hot flow itself, via synchrotron emission  (see e.g. \citealt{PV14}).
The Comptonization  produces a power-law-like X-ray spectrum with a high-energy cutoff  determined by the hot flow electron temperature.
Compared to accreting black holes (BH), the NS surface provides an additional seed photon source for Comptonization.
Consequently, as  more photons can participate in the up-scattering process, the typical equilibrium electron temperatures of NS systems are much lower than in BH systems \citep{G10, DGK07}. As the X-ray spectral slope is given by the product of the electron temperature and optical depth, NS systems also have softer X-ray spectra \citep{BGS17}.

When an X-ray burst occurs in the NS envelope, the number of soft seed photons entering the hot flow increases dramatically. 
It is therefore expected that successive interactions with these photons will cool the coronal electrons further, resulting in lower electron temperatures and softer Comptonization spectra from the persistent level.
The first hints of this effect were seen by the \textit{Rossi X-ray Timing Explorer} \xte/HEXTE in the burst light curve of  Aql X-1 \citep{MC03}, and subsequently in the \textit{RXTE}/PCA \citep{Jahoda96} light curves of several other bursters: IGR J17473--2721 \citep{CZZ12}, 4U 1636--536 \citep{JZC13}, \source\  \citep{JZC14a, JZC15}, and KS~1731--60 and 4U~1705--44 \citep{JZC14}.
By stacking more than a hundred bursts from the IBIS/ISGRI instrument on board  the {INTErnational Gamma-Ray Astrophysics Laboratory} (\inte; \citealt{WCdC03}),  \citet{KSK17} detected a clear flux drop above 40~keV in the light curves and spectra of the 4U~1728--34 hard-state bursts. 
Spectral evidence of X-ray burst-induced coronal cooling were also found in  4U~1636--536 by \citet{CZQ18}.

However, a few conflicting cases remain. The  flux deficits detected in  4U 1728--34 by \citet{KSK17} were not found in the \xte/PCA light curves by \citet{JZC14}. Additionally, \citet{itZHK99} detected an increase of the hard X-ray flux ($>$\,30~keV) in a \textit{BeppoSAX} burst from \source, while \citet{JZC14a,JZC15} detected significant hard X-ray deficits in the \xte/PCA data of the same source. 
\citet{DKC16} detected  a possible softening of the persistent power-law emission,  during a long burst of 4U~1608--52 seen by \textit{NuSTAR}, but did not find hints of a hard flux decrement. These discrepancies raised the concern that some of the reported  flux deficits may be of instrumental origin, such as  dead-time effects in the PCA instrument, when the thousands of X-ray burst photons hit the instrument each second, and thus the few high-energy photons above 40~keV may go undetected.

Moreover, significant soft excesses with respect to the hard-state persistent emission  were  measured in the 
burst spectra of  EXO~0748--676 \citep{AD06}, 
SAX J1808.4--3658 \citep{iZGM13,BJG19}, and Aql~X-1 \citep{KAB18}. These excesses suggest a temporary enhancement of the accretion rate induced by the burst, probably via Poynting-Robsertson drag of the accretion flow \citep{BE05}. 
Enhancements of the persistent emission during X-ray bursts  were also inferred by \citet{WGP2013,WGP15} fitting the  \xte/PCA spectra of a large sample of bursters, and by \citet{KBK14} during a superburst of 4U~1636--536. 

\source, the  `textbook burster' \citep{Bildsten00},  displays 
 stable bursting behaviour and  case 1 mixed H/He bursts \citep{FHM81} with quasi-periodic  recurrence over long periods of time. This regularity  also lead to the nickname `{clocked burster}' \citep{UBB99}.
 However, the burst waiting times have actually been observed to  vary between roughly 5 and 3 hours over the years, in response to variations in the mass accretion rate onto the NS \citep{CintZV03, GCK04,GMH08,TGR08}. The accretion rate onto \source\ is modest,  about 13 \% of the Eddington rate \citep{CGinZ16}. Thus, 
it  remained in the hard spectral state (or  island state; see \citealt{HvdK89}) for years, displaying only brief excursions to a softer X-ray state in 2004 and 2014 during which the accretion rate did not show significant variations (see e.g. \citealt{RJR16, CGinZ16}). Its hard-state spectrum can be nicely modelled  with a hard ($\Gamma \sim 1.5$) power law with a cutoff at about 50 keV \citep{CFP10}, and a weak high-energy tail that dominates the emission above $\sim 150$~keV \citep{RJR16}.
 In July 2015, \source\ transitioned  to a soft state, where it has remained since then  \citep{SGA18}, except for  brief hard-state episodes \citep{JSZ18}. In February 2018, the first superburst from \source\  
was detected by the Gas Slit Camera (GSC, \citealt{MTN11}) on board MAXI \citep{MMK09}. 

As the bursts until 2015 did not reach the Eddington flux \citep{itZHK99,ZCG12,CGinZ16} the burst temperatures are low enough that their contribution above 30~keV is minimal.
Therefore,  the IBIS/ISGRI \citep{ULdC03} instrument on board \inte\, being sensitive to photons only above $\sim17$~keV, it does not suffer from the  dead-time effects that can severely affect the \xte/PCA and \textsl{NuSTAR} instruments. In this
paper, we present a combined analysis of all \textsl{INTEGRAL}-detected X-ray
bursts until 2015, combined with soft-energy data in the same time frame
as measured with the RGS on board \textsl{XMM-Newton} \citep{dHBK01,dVdHGR14} to study at optimum sensitivity changes in the persistent spectrum during an X-ray burst. 
The resulting broad-band spectra provide unambiguous evidence of  X-ray burst-induced coronal cooling of \source.  Simultaneously, we find a soft X-ray excess when fitting the burst spectrum using a simple blackbody model. 


\begin{figure*}[!ht]
\begin{center}
\includegraphics[width=18.5cm]{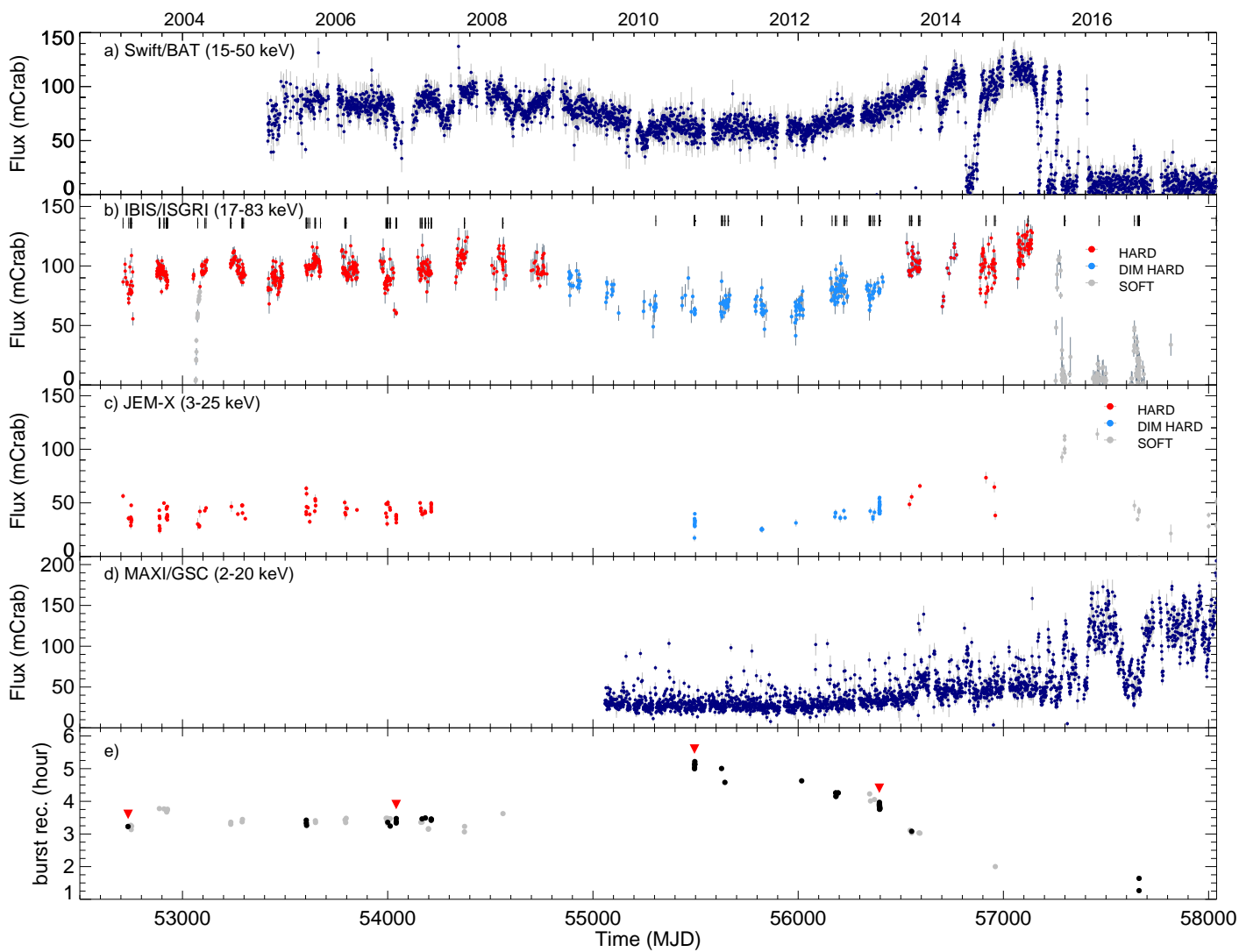}
\end{center}
\caption{Light curves of \source\ during the period analysed in this work. \textbf{Panel a)} \textit{Swift}/BAT light curve in the 15--50\,keV energy range.  \textbf{Panel b)} IBIS/ISGRI light curve in the 17--83\,keV energy range. Shown in red are the hard-state periods (MJD 52700--54800, MJD 54800--56400), and in light blue and grey,  respectively, the dim-hard periods (MJD 54800--54800) and soft-state periods (MJD 56400 onward). The small vertical lines on top of the light curve indicate the times of detection of Type I X-ray bursts  by JEM-X. To build a homogeneous sample, and avoid the systematics effects caused by variations of the persistent emission, burst stacking was performed using the hard-state data from the period MJD 52700--54800. \textbf{Panel c)} JEM-X  light curve in the 3--25\,keV energy range. \textbf{Panel d)} MAXI light curve in the 2--20\,keV energy range. \textbf{Panel e)} Recurrence times of the Type I X-ray bursts found in this work. Shown in  black are the recurrence times measured directly (see examples in Fig. \ref{fig:burst_recurrence}) and in grey  the derived recurrence times (as described in the text). The red triangles correspond to the times of detection of the four groups of bursts displayed in Fig. \ref{fig:burst_recurrence}.} 
\label{fig:overall_lcr}
\end{figure*}

\section{Observations and data reduction}
\label{sect:Obs}

\subsection{INTEGRAL}
\label{subsect:inte}
We  analysed all the available archival X-ray data of \source\ since the beginning of the \inte\ scientific operations in January 2003  until the spring of 2018, thus covering 15 years of data.
We used data from the JEM-X and IBIS/ISGRI instruments. JEM-X, consisting of two identical units X1
and X2, is sensitive in the 3--35 keV range and provides an angular resolution of $3\arcmin$, while IBIS/ISGRI is sensitive in the 15 keV to 10 MeV range, and provides an angular resolution of $12\arcmin$.
The data were reduced using standard procedures with the \inte\ Offline Science Analysis (OSA) provided by the ISDC.\footnote{ISDC Data Centre for Astrophysics, http://www.isdc.unige.ch/} The JEM-X data were analysed using osa v11.0, released in October 2018. The IBIS/ISGRI data were analysed using osa v10.2, released in December 2015. At the time of writing this paper, due to the lack of IBIS/ISGRI calibration files, the validity of OSA11.0 was  limited to data later than revolution 1626. Early in the {\sl INTEGRAL} mission, only one of the two JEM-X units was operated at a time. From revolution 976 onward, JEM-X operated with both units active. Since JEM-X1 was operational almost 14 years out of the 15 we analysed, we  based our spectral analysis on the data from JEM-X1 and used JEM-X2 data only to fill the gaps in the long-term  light curve, and derive burst recurrence times. The total JEM-X exposure accumulated in \source\  over this period is 2.17 Ms. However, we limited our spectral analysis to observations where the  angular distance between the source location and spacecraft pointing was less than 3.5 degrees, which rendered a total exposure of 954 ks. The selection criteria guaranteed that  \source\ was always within the JEM-X half-response field of view (FOV) and IBIS/ISGRI fully coded FoV (FCFOV), so the measured fluxes and hardness ratios were not affected by flux reconstruction systematic effects towards the edges of the partially coded field of view (PCFOV) that may contaminate the light curves and images. However, to minimize the gaps between successive bursts and improve the calculation of the burst recurrence times, all the available data were used in the burst searching step, as the burst detection times are not sensitive to off-axis effects. 

The X-ray bursts were identified on the 2s binned JEM-X light curves of \source\ in the 3--25~keV band extracted at the science window level (typical duration from 30~min to 1 hour). The burst identification was performed as follows: we computed the mean and standard deviation of the source count rate within each science window, and for each time bin compared the source count rate with these values. When the bin rate exceeded 6$\sigma$ of the persistent level, we identified the potential onset of an X-ray burst. The peak of the burst and subsequent burst decay were identified in the system light curve to verify that the rise and decay profile were consistent with those of a Type I X-ray burst. The burst detection was further confirmed by extracting the image of the field within a time interval restricted to the burst duration in order to discard contamination to the source light curve by an event originated in another source in the FOV.
The burst search resulted in a sample of 148 X-ray bursts.\footnote{Part of this work is also fed into a larger type I X-ray burst database, MINBAR; see http://burst.sci.monash.edu/minbar.} 

Because the hard-state burst peak fluxes and cooling timescales of \source\ are quite similar over long periods of time (years), we can stack the hard-state bursts together to obtain burst spectra of increased S/N. We identified the periods when \source\ was in the hard state and from these selected the intervals when \source\ displayed stable fluxes and bursting properties in order to minimize systematic effects due to poor characterization of the persistent emission or burst spectra. These criteria are fulfilled in the period MJD 52700--54800 ({\sl INTEGRAL} revolutions 50 to 669). In this period, the IBIS/ISGRI  (17--80\,keV) flux  was in the range 80--130\,mCrab and the JEM-X flux (3--25 keV) was 20--60\,mCrab.
Good time intervals (GTIs) were created to extract and stack five separate spectra along the burst profile, in the intervals 2--12~s, 12--32~s,  32--62~s, 62--112~s, and 112--162~s from the burst onsets.

\subsection{XMM-Newton}
\label{subsect:xmm}

\source\ was  observed by the Reflection Grating Spectrometer (RGS) on board \xmm\  in three epochs: 2002, 2003, and 2012. Because the 2002 data set  did not overlap with the \inte\ observations, we only analysed the 2003 and 2012 data sets. 
\source\ was  observed  during two consecutive revolutions between April 6 and 9, 2003 (revs. 609 and 610; 199 ks total exposure, \citealt{KMM07,iZGM13}), and in four revolutions between September 12 and 26, 2012 (revs. 2237, 2238, 2243, 2244; 139 ks total exposure). Although quasi-simultaneous with the \inte\ data set, the 2012 data were finally dropped from this analysis, as described in Section \ref{subsect:pers}.
We used the XMM-SAS version 17.0.0 to process these data. The data were filtered for high background periods using the SAS task {\texttt rgsfilter}. 
The spectra and light curves were then generated using the SAS task {\texttt rgsproc}. 

Light curves were extracted in the 0.5--2\,keV range with a time resolution of 1 s. 
The burst searching routines described in Sect. \ref{subsect:inte} were also applied to the one-second binned RGS light curve.  After filtering for high background, these resulted in the detection of 13 Type I X-ray bursts during the 2003 observations and 8 Type I X-ray bursts during the 2012 observations. The 2003 RGS data set partially overlaps with the  \inte\ observations. 
Comparing the two data sets, we identify one X-ray burst detected simultaneously in the barycentric time corrected light curves of RGS (OBS ID 0150390301) and JEM-X (Pointing ID 005900400010.001). 
This burst was used to calibrate the simultaneity of the  JEM-X and RGS burst timescales, and to verify that it was possible to build consistent RGS, JEM-X, and IBIS/ISGRI spectra stacked over the same integration periods along the burst profile (2--12\,s, 12--32\,s,  32--62\,s, 62--112 s,\, and 112--162 s\, from the burst onsets).

\section{Results}
\label{sect:Res}
\subsection{Persistent emission}
\label{subsect:pers}

The light curves of \source\ over the 15 years analysed in this work are shown in Fig. \ref{fig:overall_lcr}.  We display the IBIS/ISGRI (17--80\,keV) and JEM-X (3--25\,keV) light curves constructed from the existing \inte\ data, complemented with the publicly available daily light curves of MAXI (2--20\,keV; \citealt{MMK09}) and \swift/BAT (15--50\,keV; \citealt{Krimm2013}). Burst recurrence times derived from this analysis are also shown for reference.

\source\ remained persistently in the hard state from its discovery in 1988 until 2015,  with bolometric flux in the range 1.8--3.4\, $\times10^{-9 }\,{\rm erg\,cm^{-2}\,s^{-1}}$ \citep{TGR08}. In 2015 it transited to the soft state, except for brief excursions to the hard state \citep{JSZ18}.  
Inspecting the IBIS/ISGRI and JEMX light curves, we
identify the epochs when \source\ was detected in the hard state (MJD 52700--57200) and soft state (57200 onward).  However, despite the expected stable properties during the hard state, we note a period of reduced ($\sim$80\%)  X-ray emission in the IBIS/ISGRI and JEM-X light curves (MJD 54800--56400; blue points in Fig. \ref{fig:overall_lcr}b). Because the X-ray fluxes are not as hard in this period compared to the hard state, 
hereafter we  refer to it as {the dim hard state}.

To prevent  variations in the source persistent emission from resulting in systematic effects when stacking the burst spectra, we restricted the analysis of the hard-state burst spectra to the period  MJD 52700--54800. Although comparable IBIS/ISGRI fluxes were also measured  in the period MJD 56400--57200, the JEM-X flux gradually increased during that epoch (see Fig. \ref{fig:overall_lcr} c), which resulted in artefacts in the derived spectrum. Therefore, these data were not used to build the hard-state spectrum. 
The hard-state persistent emission spectra were properly fit with a single cutoff power-law model, modified by the interstellar absorption, which uses abundances from \cite{WAMc00}. 
When fitting the hard-state \xmm/RGS spectrum with high S/N and resolution, the interstellar absorption edges were not clearly described, as  detailed in \citet{PKC10}. Rather than performing a detailed modelling of the absorption by the interstellar medium (ISM), as in \citet{PKC10}, we instead binned the RGS data by a factor of three and added 3 \%\ systematic errors to all the RGS energy bins.
Moreover, in the spectral modelling we also added an additional edge near the neon K edge ($E_{\rm edge}=0.872_{-0.002}^{+0.008}$~keV, $\tau = 0.111 \pm 0.007$), which was needed for a good fit to the data. 
The best fitting parameters for the hard-state spectra were $N_{\rm H} = 3.36_{-0.01}^{+0.03} \times 10^{21}$~cm$^{-2}$, $\Gamma=1.491_{-0.007}^{+0.009}$, $E_{\rm cut}=57.6_{-0.5}^{+0.8}$~keV, $K_{\rm c}=0.175_{-0.002}^{+0.003}$ ($\chi^2 / {\rm d.o.f.} = 358.8/449$), in agreement with \citet{CFP10} and \citet{RJR16}.  
Our data do not require an additional Comptonization component, as used by e.g. \citet{RJR16}, when extending the spectral fits to harder energies (370\,keV). 
We also do not find  a significant contribution from the accretion disc in soft X-rays, as claimed by \citet{TGR08} and \citet{OSZ16}.
We also tested the Comptonization models \textsc{nthcomp} \citep{ZJM96,ZDS99} and \textsc{compps} \citep{PS96} available in \textsc{xspec}, but these resulted in poorer fits and wavy residuals in the IBIS/ISGRI band and  therefore were not considered further.

In the dim hard  state  the ISGRI and JEM-X light curves both showed significant flux variations over time. The combination of these averaged spectra with the September 2012 RGS spectrum results in wavy residuals (unless adding unreasonable model components), likely because at the epoch of the RGS observations the spectra were somewhat harder than the average during 2009--2013 (see Fig. \ref{fig:overall_lcr}). To compare the burst stacked spectra with the persistent emission, we need to average the persistent emission using  all the data where we detect bursts. 
Fitting the average dim-hard-state spectrum, using only  the \inte\ data we derive the following parameters $\Gamma = 1.67\pm0.03$ and  $E_{cut}=85\pm6$, ($\chi^{2}/d.o.f.=29.6/24$). 
The average dim-hard-state burst profiles resemble the hard-state ones (see e.g. the light curve shapes in Fig. 3) and after stacking the dim-hard-state burst spectra, we found qualitatively similar results to those obtained for the hard state (detailed below), but with larger errors, as the stacked dim-hard-state burst spectra are noisier. This is  because there are fewer dim-hard-state bursts to stack and the flux is lower than in the hard state. 
As our main aim is to characterize the burst-induced changes in the Comptonized component, and these are unambiguously determined using the hard-state data set, we  concentrate hereafter on the analysis of the influence of the hard-state bursts on the hard-state persistent emission spectra. 

\begin{figure}[!ht]
\includegraphics[width=9cm]{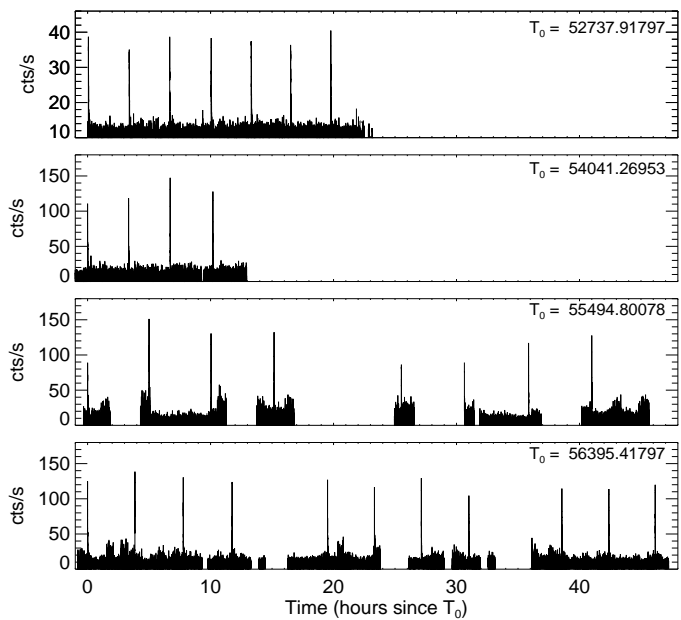}
\caption{Detailed view of four trains of bursts. The top panel shows the RGS light curve of obs ID 01050390301. The time resolution is 5 seconds (0.5--2\,keV energy range). The other panels provide JEM-X light curves built in the 3--25\,keV energy range, with a time resolution of 5\,s. For reference, the time of occurrence of these trains of bursts are marked in Fig.~\ref{fig:overall_lcr}e. The average recurrence times per panel are (from top to bottom): 3.23$\pm$0.05, 3.39$\pm$0.05\,h, 5.11$\pm$0.07\,h,  and 3.86$\pm$0.06\,h.} 
\label{fig:burst_recurrence}
\end{figure}

\subsection{Bursting behaviour}
Different bursting properties were observed in the hard, dim hard, and soft states, for which the total number of  bursts detected  by JEM-X were 90, 49, and 9, respectively. Burst waiting times were computed whenever possible using the JEM-X and RGS data sets. The derived values are displayed in Fig. \ref{fig:overall_lcr}e.  In some cases, the source was continuously in the JEM-X FOV for several hours (see Fig. \ref{fig:burst_recurrence}), and the burst recurrence times could be directly measured from the data (black points in Fig. \ref{fig:overall_lcr}e). However, gaps are frequent in the JEM-X light curves of \source\ due to the INTEGRAL dithering strategy and consecutive bursts can be missed (as happens in the two lower panels of Fig. \ref{fig:burst_recurrence}). This would result in measured burst recurrence times which are a multiple of the actual values. In such cases we estimated the burst recurrence times by correcting the measured waiting times by the duration of the gaps in the JEM-X light curves (grey points in Fig. \ref{fig:overall_lcr}e). 

The system consistently displayed quasi-periodic (clocked) burst recurrence times in the hard state (MJD 52700--54800), i.e. over a period of 5 years, when the soft and hard X-ray fluxes displayed roughly constant values.  We measure burst waiting times between 3.13 and 3.76 hours, consistent with measurements by \cite{GMH08} around this period, but significantly shorter than the average burst recurrence time   observed  between 1996 and 1998 (5.76\,h, \citealt{UBB99,GMH08}),  when the system displayed persistent hard-state bolometric fluxes in the range  1.9–2.2 $\times10^{-9 }\,{\rm erg\,cm^{-2}\,s^{-1}}$ (\citealt{UBB99,CFP10}), 40--50\% lower than the hard-state fluxes we observed  (3.8 $\times10^{-9 }\,{\rm erg\,cm^{-2}\,s^{-1}}$). 

Contrary to the stable burst waiting times  observed during the hard state, the burst recurrence times decreased monotonically  from 5.11  to 3.02 hours during the dim hard state, while the soft X-ray emission  increased (see Fig. \ref{fig:overall_lcr}). Despite the overall decreasing trend, the clocked behaviour still prevailed when looking at the light curves on short timescales (hours; see Fig. \ref{fig:burst_recurrence}).   
The detection of the decreasing burst waiting time  happened  after a long gap in the JEM-X light curve (between MJD 54300 and 55500), so we could not observe the  transition from the quasi-stable bursting regime to the decreasing burst recurrence time regime.
 The decrease in burst waiting time continued even further during the soft state, when recurrence times of 1.62 hours were measured. 

\begin{figure}
\begin{center}
\includegraphics[width=9cm]{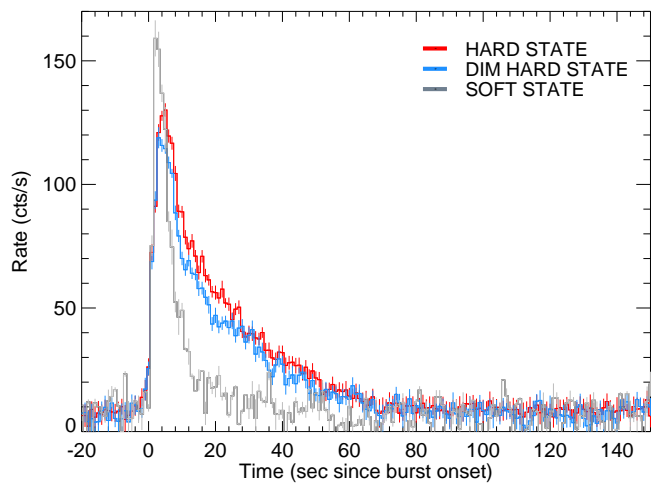}
\caption{Comparison of the JEM-X (3-25 keV) average burst profiles detected during the three spectral states described in Sect. \ref{sect:Res}: hard, dim hard, and soft.}
\end{center}
\label{fig:burst_profile_comparison}
\end{figure}

\begin{figure}[!ht]
\begin{center}
  \includegraphics[width=18.3cm,angle=90]{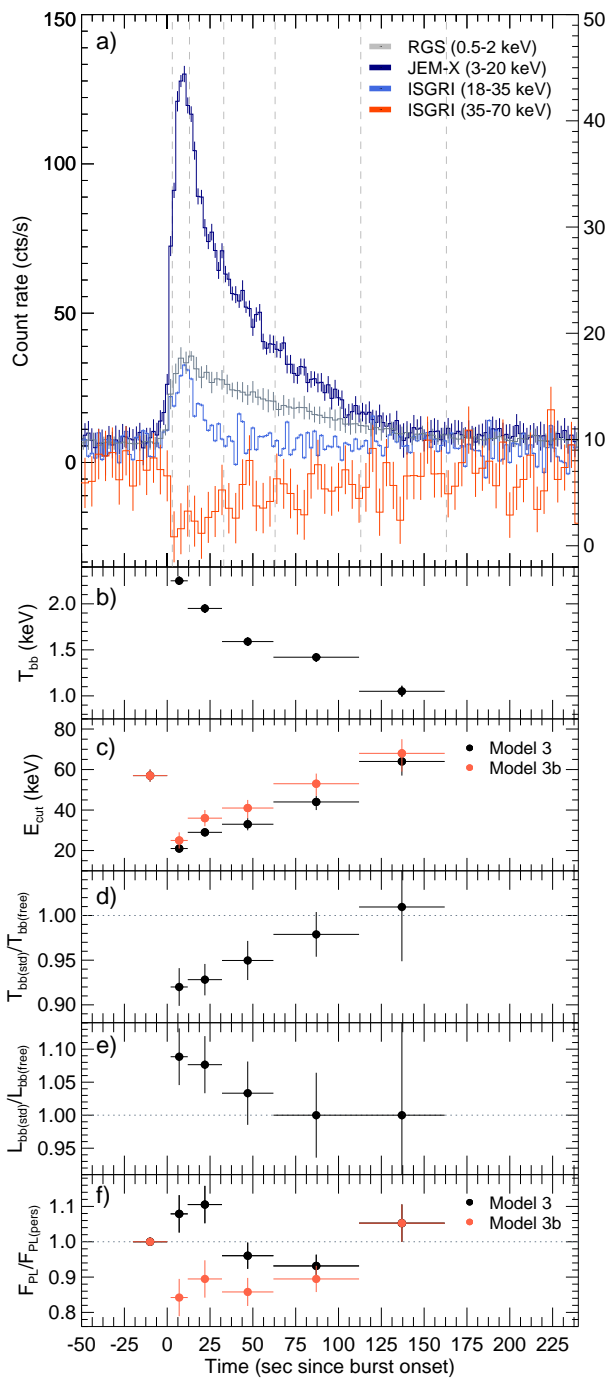}
\end{center}
\caption{{\bf Panel a)} Average burst profile in four energy bands. The RGS, JEM-X, and  IBIS/ISGRI (18--35\,keV) light curves  refer to the left vertical axis. The IBIS/ISGRI (35--70\,keV) light curve  refers to the right vertical axes. The profiles were built as described in sect. \ref{sect:Obs}. 
The dashed vertical lines indicate the intervals over which the spectra displayed in Fig. \ref{fig:ave_burst_spectra} were stacked. {\bf Panels b and c)} Evolution of the burst temperature and  cutoff energy derived from the fits to the stacked burst spectra, using Model 3 and 3b (free persistent emission parameters).  \textbf{Panels d and e)} Comparison of the burst temperatures and fluxes derived using the standard fitting procedure (fixed persistent emission parameters: Model 1) with the results derived from Model 3. A dotted line is drawn at the position that equals both values. \textbf{Panel f)} Comparison of the power-law flux derived using Models 3 and 3b with the non-bursting persistent emission flux.} 
\label{fig:burst_profile}

\end{figure}

We also built the average JEM-X burst profile in the hard, dim hard, and soft states (3--25\,keV; see Fig. 3). The burst profiles in the hard and dim hard states were very similar, although with slightly different peak count rates: (130$\pm$3\,counts s$^{-1}$) in the hard state and 119$\pm$4\,count~s$^{-1}$ in the dim hard state)\footnote{For reference, 1 Crab=250 count~s$^{-1}$ in the 3--25\,keV energy range.}. Long tails extending up to 100 seconds after the burst peak were observed in the hard and  dim hard average burst profiles.  The burst profile changed dramatically in the soft state, when peak count rates of 159$\pm$3\,count\textbf{s}~s$^{-1}$ were observed, while the burst duration decreased  substantially (to $\sim$\,15 seconds). \cite{CGinZ16} found that most of the soft state bursts of \source\ reached the Eddington luminosity. If this was also the case for the JEM-X soft state bursts averaged here,  we find that the hard and dim hard average bursts would reach peak luminosities of 74\% and 81\% $L_{\rm Edd}$, respectively by comparing their average burst peak count rates with soft state average count rate.
 
We do not observe in our data the long ($\sim$\,5000\,s) X-ray tails detected by \citet{ZKC09} in the \source\ burst profile light curves, probably due to a poorer JEM-X sensitivity  compared to  \textsl{RXTE}/PCA. We also note  that the typical duration of an \inte\ pointing is 1800--3600 seconds.

\subsection{Persistent emission changes during hard-state bursts}

We display in Fig. \ref{fig:burst_profile} the average  burst profiles of \source\  in the hard state (MJD 52700 to 54800) in four energy bands. The \inte\ burst profiles were built averaging the light curves of  48 X-ray bursts, in the JEM-X 3--20\,keV energy range (2\,s binning) and in the IBIS/ISGRI 18--35 and 35--70 keV energy range (4\,s binning).  The RGS profile was built averaging the light curves of the 13 bursts detected in the interval MJD 52736--52739. 

The JEM-X 3--20\,keV light curve displays the characteristic burst profile of \source\ in the hard state, with 8\,s rise time and a long decay (e-folding decay time was 19.6\,s). 
The burst emission is still detectable in the IBIS/ISGRI light curve in the 18--35\,keV range. However, in the 35--70\,keV energy range,  we observe a clear decrement in the emission of $\sim$\,80\%  with respect to the pre- and post-burst values.

\begin{figure*}[ht]
\begin{tabular}{lll}
\includegraphics[width=5.8cm]{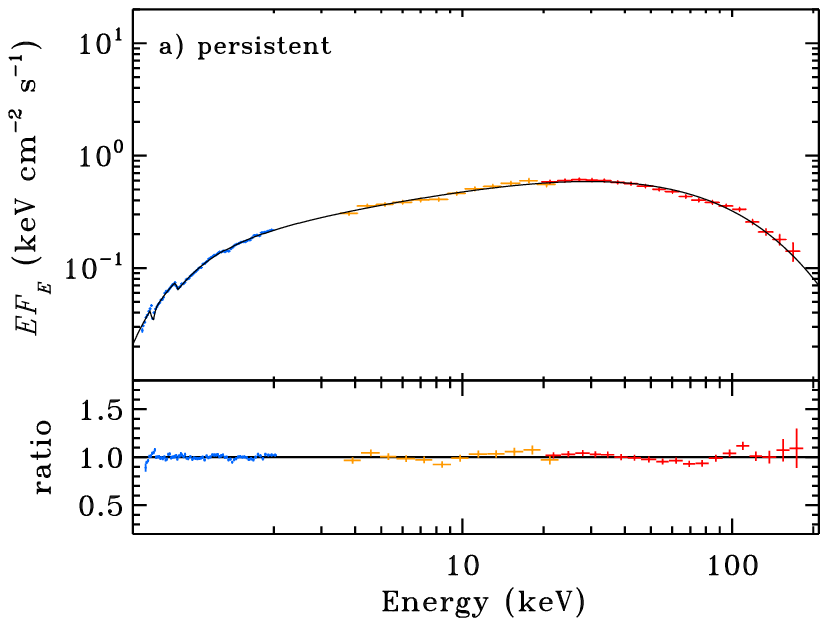} & \includegraphics[width=5.8cm]{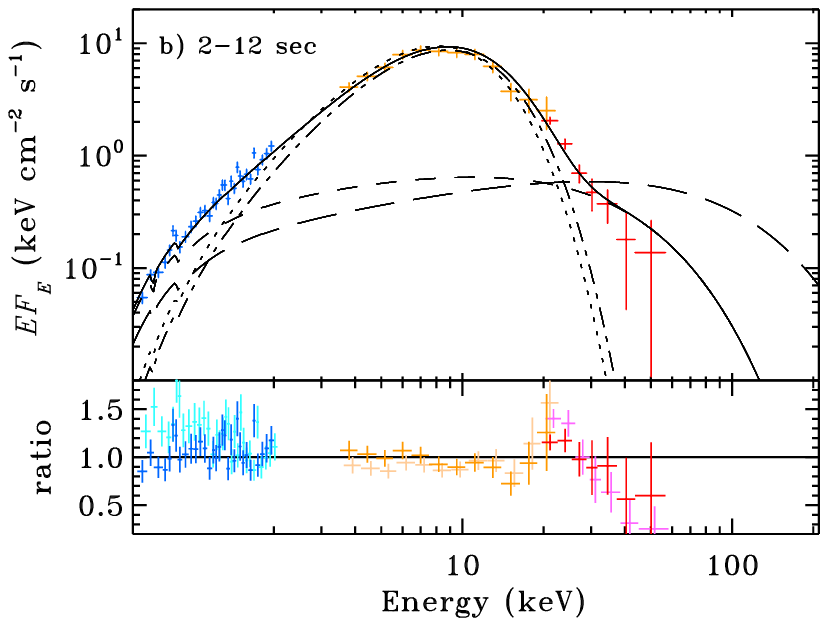} & \includegraphics[width=5.8cm]{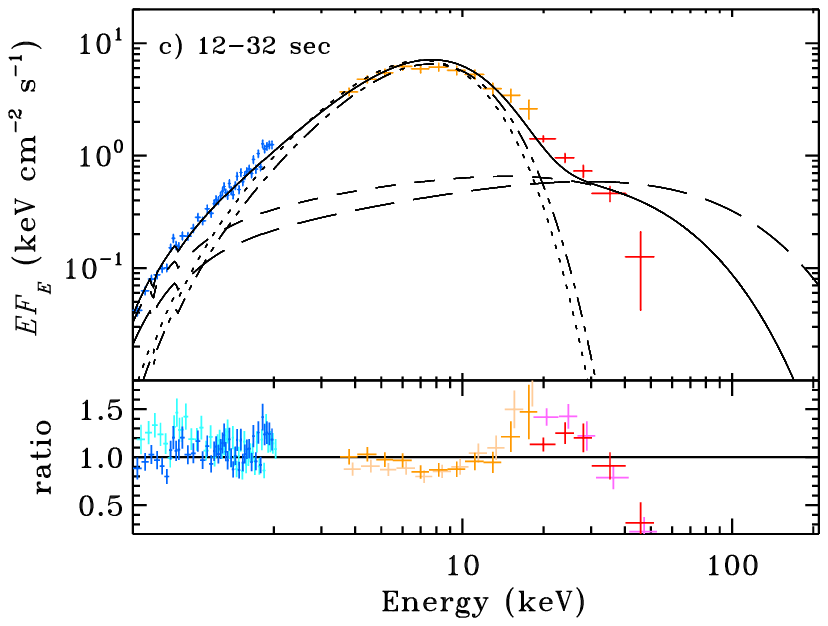} \\
\includegraphics[width=5.8cm]{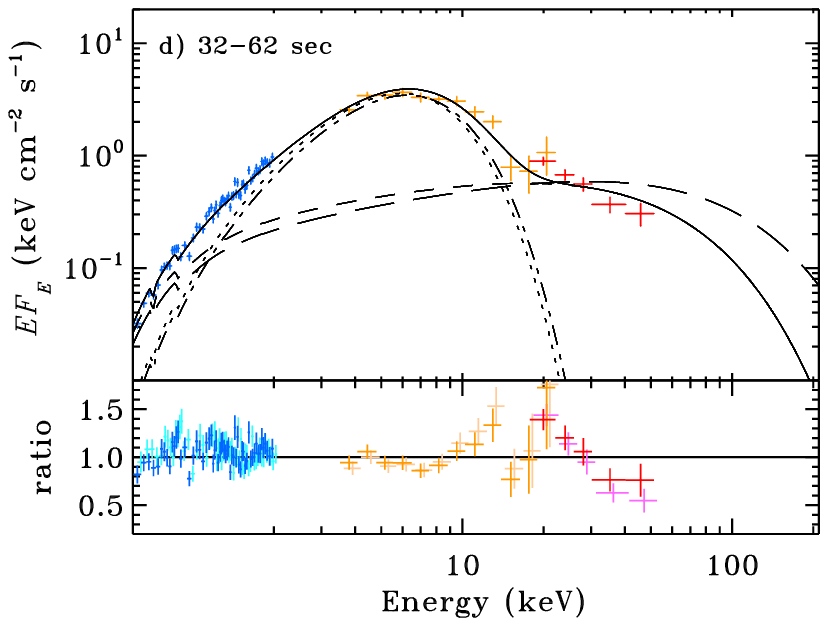} & \includegraphics[width=5.8cm]{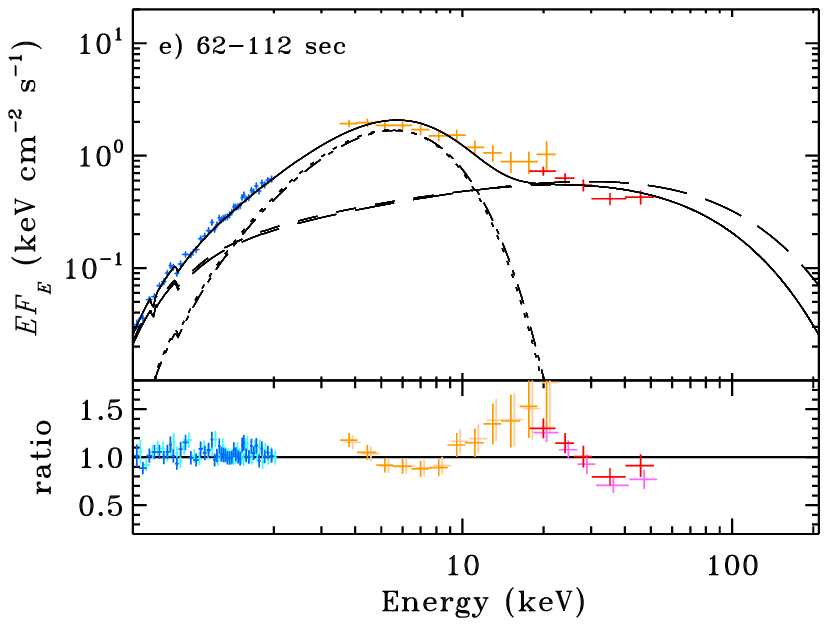} & \includegraphics[width=5.8cm]{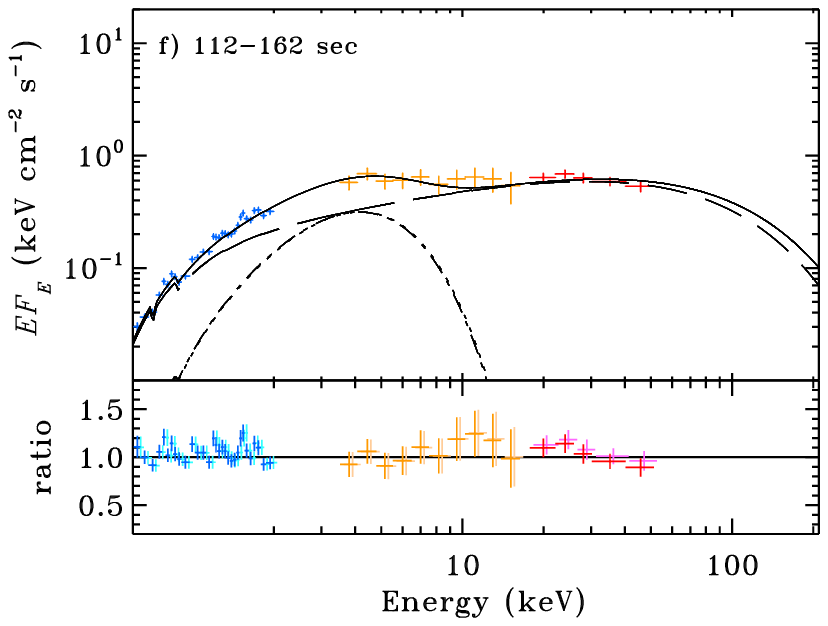}\\
\end{tabular}
\caption{\textbf{Panel a)} Average hard-state persistent emission spectrum. Shown are the RGS, JEM-X, and ISGRI spectra (in  blue, orange, and red, respectively). 
\textbf{Panels b--f)} Average hard-state burst spectra stacked over the five time intervals shown in Fig. \ref{fig:burst_profile}. The spectral parameters derived from the fits to these spectra are provided in Table \ref{tab:burstparams}.  The black continuous line shows the resulting spectra (burst+persistent emission) when fitting the data using Model 3, where the \textsc{cutoffpl} normalization and cutoff energy during the burst are treated as free parameters (short-dashed line). The dot-dashed line shows the corresponding blackbody component.  Highlighted  in the data--model panels (in cyan, brown, and pink) are the residuals in the case where the \textsc{cutoffpl} component during the burst had been fixed to the best fitting persistent emission (Model 1,  long-dashed line). The dotted  line shows the corresponding blackbody component.}
\label{fig:ave_burst_spectra}
\end{figure*}

We selected five intervals along the average burst profile (shown in Fig. \ref{fig:burst_profile}a) and for each hard-state burst we built the corresponding GTI files.  The individual burst spectra do not provide the necessary signal-to-noise ratio to perform spectral fits, but when stacking the spectra from the individual bursts we can build spectra that extend up to $\sim60$~keV. 
We show in Fig. \ref{fig:ave_burst_spectra} the persistent spectrum (panel a) and five spectra along the burst profile, spanning times between  2--12~s (panel b),   12--32~s (panel c),   32--62~s (panel d),   62--112~s (panel e), and   112--162~s (panel f) from the burst onsets. During the burst we observe a clear hard X-ray decrement, consistent with the drop observed in the burst average light curves. To account for this,
we fitted the stacked burst spectra using five different models. 
The best fitting parameters are collected in Table \ref{tab:burstparams}.
Model 1 (where  the burst spectrum is fitted with a blackbody model, assuming that the persistent emission spectrum does not change during the burst) corresponds to the standard burst fitting technique (e.g. \citealt{KHvdK02,GMH08}). 
This model can be rejected with high confidence during the burst peak and decay onset (first two intervals between 2 and 32 s) because it strongly overpredicts the hard X-ray flux and underpredicts the soft X-ray flux, resulting in a soft X-ray excess below $\sim$2\,keV. 
Model 2 (where we fit the burst spectrum using a blackbody and allow the persistent emission to vary in normalization but not in shape, as in \citealt{WGP2013,WGP15})  improves the fits significantly, but also can be rejected with high confidence in the 2--32 s intervals because it cannot simultaneously fit the observed hard X-ray deficit and the soft X-ray excess.
Using Model 2 we derive an increase in the power-law normalization of about $\approx 1.3$. This 
corresponds to the $f_\textrm{a}$-parameter in \citet{WGP2013,WGP15}.
Model 3  (where we fit the burst spectrum using a blackbody component and allow the power-law normalization and cutoff energy to vary) is the best fitting model of the three blackbody cases considered as it allows us to fit simultaneously the observed excess at soft X-rays and the decrement at hard X-rays.
For example, during the burst peak (2--12~s interval) we obtain a significant improvement of the fit ($\Delta \chi^2 = 64.7$) for two extra degrees of freedom in the model.
We  also tried to fit the persistent emission leaving the power-law index as a free parameter, but our fits failed to constrain all the parameters simultaneously.

When fitting the burst spectrum with Model 3, we observe an evolution of the cutoff energy during the burst. The cutoff energy drops from the persistent value of 57.6~keV down to 21~keV during the burst peak  (2--12~s interval), and then gradually increases to being consistent with the persistent value during the 112--162~s interval (see Table  \ref{tab:burstparams}, Fig. \ref{fig:burst_profile}b).
Additionally, we observe that the  blackbody parameters derived using Model 3 are significantly different from those derived applying the standard burst analysis (Model 1). The blackbody temperatures derived from Model 3 are systematically higher, and at the same time the blackbody luminosity is lower in Model 3 than in Model 1 (see Fig. \ref{fig:burst_profile} d, e).
The standard burst analysis (Model 1) overpredicts the blackbody luminosity in the 2--12~s interval by $9^{+5}_{-5}$ \%, while in the 32--62~s interval onward, the two methods give  similar luminosities  (see Fig. \ref{fig:burst_profile}e). 

The Model 3 results suggest that the observed soft X-ray excess could result from the softening of the Comptonized component. However, the soft excess can also be due to inaccurate modelling of the burst spectrum. Theoretical NS atmosphere models show deviations (excesses) of the burst spectra from  blackbody shapes, below $\sim$\,3\,keV \citep{SPW11,SPW12,Suleimanov2018}. Therefore, we replaced the blackbody component in our Model~1 and Model~3 by the \textsc{burstatmo} model,  based on the NS atmosphere modelling of \citet{SPW11,SPW12} and fitted again the burst spectra.
The \textsc{burstatmo} model, available in \textsc{xspec}, has five parameters: NS radius ($R_\textrm{NS}$), luminosity expressed as the ratio to the Eddington luminosity ($L/L_{\rm Edd}$), distance ($d$), logarithm of gravity ($\log g$), and chemical composition. 
The \textsc{burstatmo} normalization $K_\textsc{burstatmo}$ is related to  the anisotropy of the NS emitting region and other geometrical effects and  should be close to unity.
We assumed a  low metallicity atmosphere ($Z=0.01 \times Z_{\rm sun}$), $R_\textrm{NS} = 10.88$ km, $\log g = 14.3$ (which together give a NS mass of $M_\textrm{NS} = 1.4 \Msun$), and  $d = 5.7$ kpc \citep{CGinZ16}.

The Model~1 fits are improved significantly when we replace the blackbody component with \textsc{burstatmo} (Model~1b; see Table \ref{tab:burstparams}). In the first three intervals with the highest burst fluxes (2--12, 12--32, and 32--62 s) the \textsc{burstatmo} model results in  $\Delta \chi^2 = 65.0, 79.3, 25.5$, respectively, for  Model~1 for the same number of d.o.f.; however,  Model~1b still reproduces the observed hard X-ray flux decrement. 
The spectral fits are  improved further by letting the parameters of the cutoff power-law  component vary (Model~3b). When fitting the spectra with Model~3b, the cutoff energy  evolves during the burst as observed in Model~3, even though the best fitting values are slightly higher in Model~3b (see Fig. \ref{fig:burst_profile}c).
The \textsc{burstatmo} model normalization, $K_{\textsc{burstatmo}}$, is at a constant value for the first 62 seconds (approximately 0.69 or 0.64 for Model 1b and 3b, respectively), while later on the normalization drops. However, 
$K_{\textsc{burstatmo}}$ depends on the assumed distance, and  for  $d = 7.2$ kpc, $K_{\textsc{burstatmo}}$ tends to unity.

In Table \ref{tab:burstparams} we also show the cutoff power-law model bolometric fluxes  ($F_\textrm{per}$) during the bursts, computed using the \textsc{cflux} model in the 0.01--1000 keV range for Models 3 and 3b  (see also Fig. \ref{fig:burst_profile}e).
We note how the cutoff power-law fluxes are practically constant at the same level as outside the bursts, with $F_\textrm{per} = (3.80\pm0.02) \times 10^{-9}\,\ergcms$ (largest deviations are at the 15 \%\  level, with 3$\sigma$ from the persistent value), even though the best fitting cutoff power-law parameters and in particular the normalization can vary by up to 80 \%.
Excesses of $\sim$10\%  in the power-law flux with respect to the flux outside the burst are observed during the burst peak if the burst spectra are measured using a blackbody model, while a flux deficit is observed during the burst peak if we fit the burst spectrum using the \textsc{burstatmo} model.

We note that the biggest dissimilarities between  \textsc{burstatmo} and the standard blackbody model appear at low energies (below $\sim$2keV) where the  \textsc{burstatmo} model predicts an excess emission with respect to the standard blackbody. 
Modelling our burst spectra with the blackbody model results in a soft excess, which is fitted by an $\sim$80\% increase in the  power-law normalization    (see $K_{\rm pl}$ values in Table 1). In contrast, the \textsc{burstatmo} model naturally fits the excess, and therefore the power-law component shows a more moderate normalization increase ($\sim$30\%). Given that the cutoff energy is very similar in the two models, the hard X-ray photon decrement caused by the decrease in $E_{\rm cut}$, together with the different  normalization increase in the \textsc{burstatmo} and blackbody models,  results in the observed differences in the power-law flux.

We also note that the burst luminosity derived with the \textsc{burstatmo} model for the 2--12 s interval is 76\% of $L_{\rm Edd}$, very close to the estimate derived from the burst profile comparison in Sect. 3.2, thus supporting the validity of the model.

\begin{table*}[!th]
\centering
\caption{\label{tab:burstparams}Best fitting parameters for the burst spectra in the five intervals shown in Fig. \ref{fig:burst_profile}. The Galactic absorption column was fixed to the best fitting value of the persistent emission ($N_{\rm H} = 3.36\times 10^{21}$~cm$^{-2}$). Model 1 follows the `standard' burst fitting approach, where the persistent emission parameters are not allowed to vary during the spectral fits. In Model 2 the persistent emission shape is fixed, but its normalization is allowed to vary (i.e. the \citealt{WGP2013,WGP15} method). In Model 3 the cutoff energy and normalization are allowed to vary. In Model 3 and 3b the persistent flux $F_\textrm{per}$ (unabsorbed in the 0.01--1000 keV range) is not a fitting parameter, but is instead calculated using the \textsc{cflux} command. For the non-burst spectrum it is $3.80\pm0.02 \times 10^{-9}\,\ergcms$, and for the burst intervals it is reported in the table.
} 
\renewcommand{\arraystretch}{1.2}
\begin{tabular}{@{}l|ccccc}
\hline\hline
Parameter  & 2--12 s                    &  12--32 s    &  32--62 s             &  62--112 s &  112--162 s      \\
\hline\hline 
\multicolumn{6}{l}{Model 1: $\Gamma = 1.49$, $E_{\rm cut} = 57.6$ keV, $K_{\rm pl} = 0.175$ } \\
\hline
$T_{\rm bb}$ & $2.07_{-0.03}^{+0.03}$              & $1.81_{-0.02}^{+0.02}$       & $1.51_{-0.02}^{+0.02}$    & $1.39_{-0.02}^{+0.02}$  & $1.06_{-0.06}^{+0.06}$ \\
$L_{\rm bb}/d_{10}^2$ & $0.246_{-0.006}^{+0.006}$  & $0.183_{-0.005}^{+0.005}$       & $0.093_{-0.003}^{+0.003}$  & $0.044_{-0.002}^{+0.002}$ & $0.0083_{-0.0009}^{+0.0009}$    \\
$\chi^2 / {\rm d.o.f.}$ & $ 193.6/114 $          & $ 370.1/253 $    & $ 344.9/323 $       & $ 414.5/413 $ & $354.6/318$ \\
\hline\hline
\multicolumn{6}{l}{Model 1b: $\Gamma = 1.49$, $E_{\rm cut} = 57.6$ keV, $K_{\rm pl} = 0.175$, $R_\textrm{NS} = 10.88$ km, $\log g = 14.3$} \\
\hline

$L/L_{\rm Edd}$ & $0.70_{-0.02}^{+0.02}$  & $0.52_{-0.02}^{+0.02}$       & $0.28_{-0.02}^{+0.02}$  & $0.196_{-0.014}^{+0.015}$ & $0.057_{-0.013}^{+0.02}$    \\
$K_\textsc{burstatmo}$ & $0.69_{-0.02}^{+0.02}$              & $0.70_{-0.02}^{+0.02}$       & $0.67_{-0.03}^{+0.03}$    & $0.46_{-0.02}^{+0.02}$  & $ 0.30_{-0.05}^{+0.06}$ \\
$\chi^2 / {\rm d.o.f.}$ & $  128.6/114  $          & $ 290.8/253  $    & $ 319.4/323 $       & $ 425.2/413 $ & $354.8/318$ \\
\hline\hline

\multicolumn{6}{l}{Model 2: $\Gamma = 1.49$, $E_{\rm cut} = 57.6$ keV  } \\
\hline
$T_{\rm bb}$ & $2.09_{-0.03}^{+0.03}$         & $1.85_{-0.02}^{+0.02}$       & $1.52_{-0.02}^{+0.02}$    & $1.39_{-0.02}^{+0.02}$ & $1.07_{-0.06}^{+0.06}$ \\
$L_{\rm bb}/d_{10}^2$ & $0.230_{-0.007}^{+0.007}$         & $0.173_{-0.005}^{+0.005}$       & $0.092_{-0.003}^{+0.003}$  & $0.044_{-0.002}^{+0.002}$ & $0.0081_{-0.0009}^{+0.0009}$   \\
$K_{\rm pl}$ & $0.229_{-0.011}^{+0.011}$         & $0.223_{-0.008}^{+0.008}$       & $0.188_{-0.006}^{+0.006}$  & $0.175_{-0.005}^{+0.005}$ & $0.178_{-0.005}^{+0.005}$    \\
$\chi^2 / {\rm d.o.f.}$ & $ 170.2/113 $          & $ 332.8/252 $    & $ 340.6/322 $       & $ 414.5/412 $ & $354.3/317$ \\
\hline\hline
\multicolumn{6}{l}{Model 3: $\Gamma = 1.49$  } \\
\hline
$T_{\rm bb}$ & $2.25_{-0.04}^{+0.04}$         & $1.95_{-0.03}^{+0.03}$       & $1.59_{-0.03}^{+0.03}$    & $1.42_{-0.03}^{+0.03}$ & $1.05_{-0.06}^{+0.06}$ \\
$L_{\rm bb}/d_{10}^2$ & $0.226_{-0.007}^{+0.007}$         & $0.170_{-0.005}^{+0.005}$       & $0.090_{-0.003}^{+0.003}$  & $0.044_{-0.002}^{+0.002}$ & $0.0083_{-0.0010}^{+0.0010}$    \\
$K_{\rm pl}$ & $0.32_{-0.02}^{+0.02}$         & $0.278_{-0.011}^{+0.011}$       & $0.223_{-0.009}^{+0.009}$  & $0.187_{-0.006}^{+0.006}$ & $0.174_{-0.006}^{+0.006}$    \\
$E_{\rm cut}$ & $21_{-2}^{+3}$         & $29_{-2}^{+2}$       & $33_{-3}^{+3}$  & $44_{-3}^{+4}$ & $64_{-6}^{+7}$     \\
$F_{\rm per}$ & $4.1_{-0.2}^{+0.2}$         & $4.2_{-0.2}^{+0.2}$       & $3.65_{-0.14}^{+0.14}$  & $3.54_{-0.12}^{+0.12}$ & $4.0_{-0.2}^{+0.2}$     \\
$\chi^2 / {\rm d.o.f.}$ & $ 105.5/112 $          & $ 279.4/251 $    & $306.9/321 $       & $ 405.6/411 $ & $353.1/316$ \\
\hline\hline
\multicolumn{6}{l}{Model 3b: $\Gamma = 1.49$, $R_\textrm{NS} = 10.88$ km, $\log g = 14.3$ } \\
\hline
$L/L_{\rm Edd}$ & $0.76_{-0.02}^{+0.02}$  & $0.56_{-0.02}^{+0.02}$       & $0.29_{-0.02}^{+0.02}$  & $0.188_{-0.015}^{+0.015}$ & $0.049_{-0.012}^{+0.015}$    \\
$K_{\rm burstatmo}$ & $0.63_{-0.03}^{+0.03}$              & $0.64_{-0.03}^{+0.03}$       & $0.65_{-0.04}^{+0.04}$    & $0.50_{-0.03}^{+0.04}$  & $0.36_{-0.08}^{+0.09}$ \\
$K_{\rm pl}$ & $0.23_{-0.02}^{+0.02}$         & $0.203_{-0.012}^{+0.012}$       & $0.179_{-0.009}^{+0.009}$  & $0.165_{-0.006}^{+0.006}$ & $0.168_{-0.006}^{+0.006}$    \\
$E_{\rm cut}$ & $25_{-4}^{+4}$         & $36_{-4}^{+4}$       & $41_{-4}^{+5}$  & $53_{-5}^{+6}$ & $68_{-7}^{+9}$     \\
$F_{\rm per}$ & $3.2_{-0.2}^{+0.2}$         & $3.4_{-0.2}^{+0.2}$       & $3.26_{-0.15}^{+0.15}$  & $3.40_{-0.13}^{+0.14}$ & $4.0_{-0.2}^{+0.2}$     \\
$\chi^2 / {\rm d.o.f.}$ & $  107.0/112  $          & $ 277.6/251  $    & $ 307.0/321 $       & $ 415.4/411 $ & $352.3/316$ \\
\hline\hline
\end{tabular}
\renewcommand{\arraystretch}{1}
\end{table*}

\section{Discussion}

The average hard-state burst profiles in four energy bands, selected in the range 0.5--70\,keV (see Fig. \ref{fig:burst_profile}), nicely illustrate the burst-induced coronal cooling effect. 
While the burst light curves in the 0.5--20\,keV energy range (RGS and JEM-X) display the characteristic burst profile (fast rise and exponential decay), and the residual burst emission still contributes to the IBIS/ISGRI 18--35\,keV  light curve,  
in the hardest band analysed here (35--70\,keV) we observe an abrupt decrement of the hard X-ray emission simultaneous with the burst peak in soft X-rays. The hard X-ray  decrement is followed by a gradual rise to pre-burst values simultaneously with the emission decay in the softer bands. The observed depletion of  hard X-ray photons is consistent with previous results by \citet{JZC14a,JZC15}, who detected significant hard X-ray deficits in the \xte/PCA light curves of \source. Our work thus confirms that the results by these authors are not of instrumental origin in the \xte/PCA. 
By stacking the burst spectra along the burst profile we were able to perform time-resolved spectroscopy in the 0.7--60\,keV energy range (see Table  \ref{tab:burstparams}, Fig. \ref{fig:ave_burst_spectra}).  The extended energy range, compared to previous works (e.g. \citealt{UBB99,CintZV03,GMH08}), allows the simultaneous observation of the burst and persistent emission, and enables the unambiguous characterization of the burst and persistent emission spectral parameters. 
The resulting spectra confirm the decrement observed in the system 35--70\,keV light curve. Simulations of the burst-induced cooling of the corona by \citet{DBB18} based on the \textsc{EQPAIR} model of \citet{Coppi92} show a gradual electron cooling in response to the burst increasing luminosity, until eventually the hard X-ray emission is suppressed.
Our observed decrease in the cutoff energy from about 60 keV down to 21 keV during the burst peak, and the gradual return of the cutoff energy back to its original value in the burst tail resemble the results of these simulations above  $\sim$2\,keV (see Fig. 10 in \citealt{DBB18}). 
Moreover, we observe that the decrement in hard X-rays is accompanied by a soft X-ray excess (below 2\,keV) not directly evident in the burst light curve alone. 
Enhancements of the persistent emission (below 3\,keV)  during X-ray bursts were previously observed in EXO~0748--676 \citep{AD06},    SAX~J1808.4--3658 \citep{iZGM13} and Aql~X-1 \citep{KAB18}. \source\ is the first case where both a hard ($>3$0\,keV) flux decrement and an enhancement of the soft X-ray flux are observed simultaneously.

Our spectral analysis does not agree with the traditional assumption that the persistent, Comptonized, emission does not vary during an X-ray burst (\citealt{vPL86,L93,KHvK02,GMH08}; Model 1 in Table  \ref{tab:burstparams}). It also shows that the assumption that the spectral shape of the persistent emission is constant during an X-ray burst and only its normalization varies (\citealt{WGP2013,WGP15}; Model 2 in Table  \ref{tab:burstparams}) does not provide a proper fit to the data  when a broader energy range is used. 

The best fits are actually provided by a model where both the persistent emission cutoff energy and normalization are allowed to vary freely (Model 3 and 3b in Table  \ref{tab:burstparams}). 
In this case, even if we saw a 30--80 \%\  increase of the persistent emission normalization with respect to the pre-burst values  (Model 3 and 3b), we actually only find a modest 10 \%\ variation of the persistent emission flux, which suggests that in \source\ the burst does not  substantially influence the mass accretion rate. 
When comparing models 1 to 1b, it is remarkable how much better the NS atmosphere model fits the soft X-ray data compared to the simple blackbody model. This suggests that in \source\ the soft X-ray excess is not related to the persistent spectrum changes, as in Model 1 the persistent emission is fixed to the pre-burst value. Moreover, when replacing the blackbody component with the atmosphere model, the persistent flux in fact seems to decrease with respect to pre-burst values.

Using NICER observations, with similar soft X-ray coverage, of an Aql X--1 hard-state non-photospheric radius expansion (PRE) burst, \citet{KAB18} found an enhancement of the persistent emission a factor of 2.5 times the pre-burst values. 
It is possible that this result is in part due the lack of hard X-ray coverage of this burst, which allowed them to assume that the persistent emission shape did not change during the burst. However,  the derived persistent emission enhancement might also be less significant if the burst spectrum was fitted with an atmosphere model instead of a blackbody.

Additionally, our results have  important implications due to the textbook  reputation of \source\ \citep{Bildsten00}. It is the target for making model--theory comparisons, and the inferred variations in the persistent emission during Type I X-ray bursts, previously not accounted for, now suggest  that the results of these model comparisons may not be accurate.
Our broad-band spectral fits have revealed biases in the determination of the average \source\ burst spectral parameters ($L_{\rm bb}, T_{\rm bb}$) when the spectra are fitted following the standard fitting approach (Model 1 in Table  \ref{tab:burstparams}). The bias is  more dramatic during the burst peak ($L_{\rm bb}$ is overestimated by $9^{+5}_{-5}$\% and $T_{\rm bb}$ is underestimated by $8^{+2}_{-2}$\%). 
The discrepancies decrease as the burst decays, being negligible  62 seconds after the burst peak (see Fig. \ref{fig:burst_profile}). In other words,  additionally, the burst-disc-corona interaction generates a much stronger luminosity bias in the peak than in the tail, distorting the burst profile from theoretical expectations. 

For example, the burst light curves of \source\ have been used in the multi-zone KEPLER light curve model \citep{WZW78,WHC04,HCG07} to calibrate the dependences between the accretion rate and burst recurrence times, burst energies, and $\alpha$-values \footnote{$\alpha$ is
 defined as the ratio of the time-averaged accretion luminosity to the time-averaged burst luminosity. It can be used to infer the composition of the burning fuel \citep{Bildsten00}}. The model calibration strongly depends on the burst peak luminosities (where we observe larger luminosity biases) as  the model parameters  are usually selected to bring the observed and predicted peak luminosities into agreement. 
Actually, large discrepancies are detected between the predicted and observed light curves during the burst rise and tails ($>$30\,s) when only the peak fluxes are used to normalize the model  \citep{HCG07,ZCG12}, while the discrepancies are smaller if the entire burst light curve is fitted \citep{ZCG12}. 
\source\ was also used by \citet{CAH16}  and \citet{MMM19}  as a calibration target to study the influence of key nuclear reaction rates on the observed burst profiles derived from the codes KEPLER \citep{HCG07}  and MESA \citep{PBD11,PCA13,PMS15,PSB18}.  The burst peak luminosity bias derived in our work 
is comparable to increasing the key $^{15}\textrm{O}(\alpha,\gamma)^{19}\textrm{Ne}$ alpha capture rate (responsible for the rp-process burning breakout from the hot CNO cycle) by a factor of 100 (see \citealt{CAH16}, their Fig. 6). 

The burst peak luminosity bias has additional  implications on the NS mass-radius constraints, used in the determination of the equation of state (EOS) of supranuclear matter \citep{LP07}. Neutron star mass and radius are usually constrained from PRE bursts (\citealt{LvPT93}). It is assumed that in PRE bursts the luminosity   at the touchdown point is equal to the Eddington luminosity. As the normalization of the burst spectrum is related to the emitting area, observations of PRE bursts can be used to put constraints on the NS mass-radius relation.
If the flux of the PRE bursts (much brighter than the sub-Eddington \source\ bursts analysed here) is also overestimated, due to the observed luminosity bias, the  NS mass-radius constraints provided by this analysis will be inaccurate.

Alternative  NS mass-radius constraints can be derived (for PRE and non-PRE bursts) by comparing the cooling tracks of  Type I X-ray bursts with theoretical  atmosphere models \citep{SP06,ZCG12,PNK14,KNL14,NSK16,NMS17}. 
The theoretical atmosphere models in general agree closely with the data, but some residuals are detected in these fits, especially when the flux drops below a certain value \citep{ZCG12,NMS17,SKM17}. Although heating of the NS atmosphere by the accreted material could be the reason for these discrepancies \citep{Suleimanov2018}, an additional explanation for these residuals 
could be the incorrect modelling of the system persistent emission, which is assumed invariant during the burst, and is tied to its pre-burst values. 
We note that  \citet{NMS17} report flux relative errors of 1–5\% in the fits to  4U~1702--429 hard-state burst spectra, consistent with the flux biases determined in our work. 

On the other hand, the burst-induced  persistent emission variation should be proportional to the persistent emission level, which is set by the mass accretion rate.
In \source\ the mass accretion rate ($\dot{M}\sim0.13 \dot{M}_{\rm Edd}$) is substantially higher than that of 4U~1702--429 and other atoll sources showing hard-state PRE bursts ($\dot{M}\sim0.01 \dot{M}_{\rm Edd}$), and therefore the burst flux bias may not be as significant in these low mass accretion rate sources as it is in \source.

\section{Summary and conclusions}
We have studied the influence of the X-ray burst photons on the hard-state persistent emission of \source\ using data collected over the period 2003--2008. 
Time-resolved spectral fits to the burst and persistent emission show an overall change in the persistent emission spectrum during the burst peak, characterized by a simultaneous hard X-ray ($>$30\,keV) decrement during the burst peak, and an enhancement of the soft X-rays ($<$2\,keV) with respect to a simple blackbody model.
The hard X-ray decrement can be explained by burst-induced cooling of the corona, which  does explain the soft X-ray excess. Instead, the soft X-ray excess largely disappears if the burst spectra are fitted with a NS atmosphere model, rather than a pure blackbody.
Other mechanisms (such as a temporary enhancement of the mass accretion rate onto the NS due to the partial collapse of the hot flow, or Poynting-Robertson drag of the accretion flow  by the burst X-ray photons) are not needed to explain the observed soft X-ray excess as we do not find significant  changes in the persistent emission bolometric flux during the bursts.

Our results show that if the burst-induced spectral changes are not taken into account when fitting the burst spectra, the burst luminosity can be overestimated by up to 9 \%\ at the peak of the \source\ bursts, while at around 60 seconds after the peak the luminosity bias becomes negligible.
That means that the interaction between the burst and the accretion flow causes a stronger luminosity bias in the peak than in the tail; therefore, it additionally distorts the burst bolometric  light curve, which is used to calibrate  theoretical burst models.
The accurate  determination of the persistent emission for the  existing burst data sets is not trivial, as most of them came from instruments operating in a reduced energy range compared to the one used in this work. 
It is thus not obvious how to correct these bursts for the burst peak luminosity bias found in this work.

\begin{acknowledgements}

The authors would like to thank J. J. M. in 't Zand for useful discussions and valuable comments during the elaboration of this work. We also thank the anonymous referee for the suggestions that helped to improve this manuscript. 
JJEK acknowledges support from the Academy of Finland grant 295114. JP and VFS were supported by the grant 14.W03.31.0021 of the Ministry of Science and Higher Education of the Russian Federation. The work of V.S. was supported by Deutsche Forschungsgemeinschaft (DFG, (grant WE 1312/51-1). Based on observations with INTEGRAL, an ESA project with instruments and science data centre funded by ESA member states (especially the PI countries: Denmark, France, Germany, Italy, Switzerland, Spain) and with the participation of Russia and the USA. Based on observations obtained with XMM-Newton, an ESA science mission with instruments and contributions directly funded by ESA Member States and NASA. This research has made use of the MAXI data provided by RIKEN, JAXA and the MAXI team.  

\end{acknowledgements}


\bibliographystyle{aa}
\bibliography{allbib.bib} 

\end{document}